\documentclass[aps,prl,showpacs,twocolumn,superscriptaddress]{revtex4}
\usepackage{amsmath}
\usepackage{amssymb}
\usepackage{graphicx}
\usepackage{verbatim}
\begin{document}
\title{Exchange bias driven by the Dzyaloshinskii-Moriya interaction and ferroelectric 
polarization at G-type antiferromagnetic perovskite interfaces}
\author{Shuai Dong}
\affiliation{Department of Physics and Astronomy, University of Tennessee, Knoxville, Tennessee 37996, USA}
\affiliation{Materials Science and Technology Division, Oak Ridge National Laboratory, Oak Ridge, Tennessee 32831, USA}
\affiliation{Nanjing National Laboratory of Microstructures, Nanjing University, Nanjing 210093, China}
\author{Kunihiko Yamauchi}
\affiliation{Consiglio Nazionale delle Ricerche-Istituto Nazionale per la Fisica della Materia (CNR-INFM), CASTI Regional Laboratory, 67100 L\textquoteright Aquila, Italy}
\author{Seiji Yunoki}
\affiliation{Computational Condensed Matter Physics Laboratory, RIKEN, Wako, Saitama 351-0198, Japan}
\affiliation{CREST, Japan Science and Technology Agency (JST), Kawaguchi, Saitama 332-0012, Japan}
\author{Rong Yu}
\affiliation{Department of Physics and Astronomy, University of Tennessee, Knoxville, Tennessee 37996, USA}
\affiliation{Materials Science and Technology Division, Oak Ridge National Laboratory, Oak Ridge, Tennessee 32831, USA}
\author{Shuhua Liang}
\affiliation{Department of Physics and Astronomy, University of Tennessee, Knoxville, Tennessee 37996, USA}
\affiliation{Materials Science and Technology Division, Oak Ridge National Laboratory, Oak Ridge, Tennessee 32831, USA}
\author{Adriana Moreo}
\affiliation{Department of Physics and Astronomy, University of Tennessee, Knoxville, Tennessee 37996, USA}
\affiliation{Materials Science and Technology Division, Oak Ridge National Laboratory, Oak Ridge, Tennessee 32831, USA}
\author{J.-M. Liu}
\affiliation{Nanjing National Laboratory of Microstructures, Nanjing University, Nanjing 210093, China}
\author{Silvia Picozzi}
\affiliation{Consiglio Nazionale delle Ricerche-Istituto Nazionale per la Fisica della Materia (CNR-INFM), CASTI Regional Laboratory, 67100 L\textquoteright Aquila, Italy}
\author{Elbio Dagotto}
\affiliation{Department of Physics and Astronomy, University of Tennessee, Knoxville, Tennessee 37996, USA}
\affiliation{Materials Science and Technology Division, Oak Ridge National Laboratory, Oak Ridge, Tennessee 32831, USA}
\date{\today}

\begin{abstract}
Exchange bias is usually rationalized invoking spin pinning effects caused by uncompensated antiferromagnetic interfaces. However, for compensated antiferromagnets other extrinsic factors, such as interface roughness or spin canting, have to be considered to produce a small uncompensation. As an alternative, here we propose two (related) possible mechanisms, driven by the intrinsic Dzyaloshinskii-Moriya interaction and ferroelectric polarization, for the explanation of exchange bias effects in perovskites with compensated G-type antiferromagnetism. One of the mechanisms is only active when a multiferroic material is involved and it is controllable by electric fields.
\end{abstract}
\pacs{75.70.Cn, 75.10.Hk, 75.30.Et, 75.80.+q}
\maketitle

\textit{Introduction.} The exchange bias (EB) effect, characterized by a shift of the magnetic hysteresis loops away from the center of symmetry at zero magnetic field, is widely reported to exist in magnetic systems where there is an interface between antiferromagnetic (AFM) and ferromagnetic (FM) (or ferrimagnetic) materials \cite{Nogues:Jmmm}. 

%This EB effect is not only conceptually important from a fundamental-science perspective, but also has crucial technological impact since it has been extensively used in magnetic devices, such as sensors and valves.

Theoretically, the EB is understood as induced by spin pinning effects at the FM/AFM  interface. An uncompensated AFM interface is usually invoked to illustrate how the pinning may work.
%Since the total magnetic moment of AFM side is nearly zero and the external magnetic filed is usually much weaker than the internal AFM exchange coupling, the spins of AFM side are almost unaffected by applied magnetic field. In contrast, the spins of FM side, which have a finite magnetic moment, can be easily redirected by magnetic field. Therefore, t
These uncompensated AFM spins at the interface are expected to pin the nearest-neighbor (NN) FM spins via the exchange coupling, giving rise to a preferred direction for the FM moments. However, despite its physical appeal, this simple picture is not enough to fully understand several real EB cases in a variety of magnetic systems. This approach usually predicts an EB larger than measured, and also fails to answer why there is EB in some fully compensated AFM interfaces \cite{Kiwi:Jmmm}.

Precisely for the subtle case of compensated AFM interfaces, extrinsic factors are also often considered, such as interface roughness \cite{Malozemoff:Prb}. Spin-canting near the interface can also contribute to the EB \cite{Koon:Prl}. 
%, such as the ``spin-flop'' at the (110) interface of body-centered AFM materials 
Other models have also been proposed, such as frozen interfacial and domain pinning. Most of these models still need a small ``frozen'' uncompensation of the AFM moments near the interface, thus remaining under much debate \cite{Kiwi:Jmmm}.

%Besides metals, the EB effect can also occur at the interfaces between FM/AFM oxides.
Recently, remarkable improvements in oxide thin-film techniques have allowed for the growth and characterization of complex oxide heterostructures with (near) atomic precision, opening an avenue for the fabrication of multifunctional devices using strongly correlated electronic materials \cite{Dagotto:Sci}. In this context, EB has been observed in BiFeO$_3$ (BFO) based heterostructures \cite{Chu:Nm}. More interestingly, the EB in multiferroic heterostructures is widely believed to be controllable by electric fields. In addition, the EB has also been observed in SrRuO$_3$/SrMnO$_3$ (SRO/SMO) superlattices \cite{Choi:Apl}. Considering that both BFO and SMO are well-known compensated G-type AFM materials (all NN spins are antiparallel) and that the interfaces are very smooth, the origin of the EB in these heterostructures remains a puzzle. The purely magnetic interactions framework stemming from traditional metallic magnetism appears incomplete to deal with the complex physics unveiled in these strongly correlated electronic systems, and to address the practical matter of how to control the EB by electric fields when a multiferroic material is involved. Therefore, new mechanisms that emphasize the many simultaneously active degrees of freedom in correlated electron systems are needed to better understand these interesting effects.

\textit{The model.} Here, we propose two (related) mechanisms for EB generation in interfaces involving FM/G-AFM perovskites. In these mechanisms, the G-AFM interface can be \emph{fully compensated}, namely the tiny uncompensation caused by various uncertain factors is no longer essential (although it can still exist). Therefore, our proposed mechanism is conceptually different from ideas based on tiny frozen uncompensated AFM moments \cite{Chu:Nm,Choi:Apl}. Instead, the interactions between spins and {\it lattice distortions} become the key intrinsic driving force for the mechanisms presented below.

Let us start with the spin-spin interaction in perovskites, with a Hamiltonian:
\begin{equation}
H=\sum_{<ij>}[J_{i,j}\vec{S}_{i}\cdot\vec{S}_{j}+\vec{D}_{i,j}\cdot(\vec{S}_{i}\times\vec{S}_{j})],
\end{equation}
where $J_{i,j}$ is the standard superexchange (SE) coupling between NN spins; $i$ and $j$ are site indices; and $\vec{S}$ are spin vectors. For several large-spin transition metal cations in perovskites, such as Mn$^{3+}$ and Fe$^{3+}$, adopting the widely-used classical approximation is reasonable. In the following, the normalization $|\vec{S}|=1$ will be used (the actual magnitude $S$ of the spins can be absorbed in a redefinition of couplings). The second term is the Dzyaloshinskii-Moriya (DM) interaction, which arises from the spin-orbit coupling \cite{Dzyaloshinsky:Jpcs,Moriya:Pr}. Since $|\vec{D}|$ is much smaller (by 
two or three orders of magnitude) than $J$ \cite{Moriya:Pr}, the DM interaction is often neglected. Originally, the DM interaction was introduced to explain the presence of weak ferromagnetism in AFM materials because the DM term can produce a small spin-canting. Recently, the DM interaction has also been highlighted as the origin of a finite ferroelectric (FE) polarization ($\vec{P}$) in multiferroic materials with spiral spin order \cite{Sergienko:Prb}.

In perovskites, the DM interaction is determined by the oxygen octahedron tilting. Usually, the A-site cations in perovskites are too small to maintain a stable cubic lattice. Then, the oxygen octahedra surrounding the B-site cations will tilt for a closer packing \cite{Woodward:Acb}. The tilting can be characterized by the Glazer notation: e.g. $a^-b^-c^+$ where the three letters denote the rotation angle amplitudes about the [100], [010], and [001] axes, respectively; the positive (negative) superscript indicates that the rotations of two neighboring octahedra, along the tilting axis, are in the same (opposite) direction \cite{Glazer:Acb}. For instance, in the orthorhombic lattices (e.g. bulk LaMnO$_3$ at low temperature ($T$)), the tilting $a^-a^-b^+$ receives the name ``GdFeO$_3$-type distortion'' and it corresponds to rotations around the [110] (dominant) and [001] (subdominant) axes of the cubic unit cell \cite{Zhou:Prb}. For the M-O-M bond (M: B-site metal and O: oxygen), this octahedral tilting moves the oxygen anion perpendicularly away from the midpoint between NN metal cations, as shown in Fig.~1(a). Since the tilting rotation is collective, the NN oxygens in the same direction (O$_1$ and O$_2$ in M-O$_1$-M-O$_2$-M) should move in opposite directions, namely the NN displacements are staggered.

\textit{DM driven EB.} From symmetry argumentations, the $\vec{D}_{i,j}$ vector should be perpendicular to the M$_i$-O-M$_j$ bond \cite{Moriya:Pr}, as shown in Fig.~1(a). Thus, the $\vec{D}$ vectors between NN bonds along the same direction are also staggered, namely $\vec{D}_{i,i+1}$=$-\vec{D}_{i+1,i+2}$. To simplify the discussion, let us consider the case where the rotations of NN octahedra along the [100] and [010] axes are in opposite directions, namely $a^-b^-c^*$ ($*$ can be $+$, $-$, or $0$).

\begin{figure}
\centerline{\includegraphics[width=0.48\textwidth]{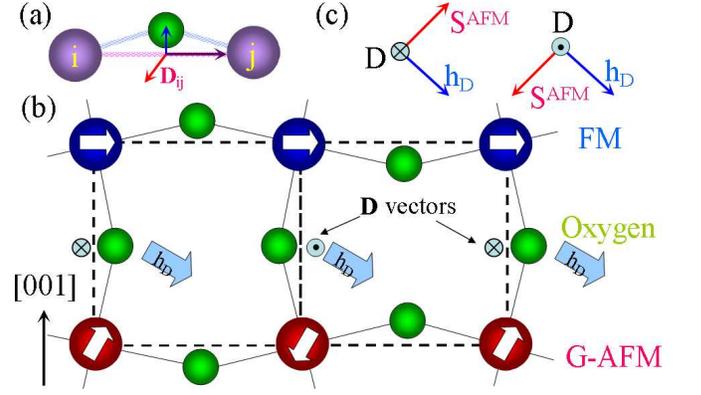}}
\caption{(Color online) (a) The (mutually perpendicular) relationship between the M$_i$-O-M$_j$ bond, oxygen displacement, and $\vec{D}_{i,j}$ vector.  (b) Sketch of the interface between FM and G-AFM perovskites, including the oxygen octahedral tilting. The staggered directions of the $\vec{D}_{ij}$ vectors at the interface are marked as in- and out-arrows, while the uniform $\vec{h}_{\rm D}$ vectors are also shown near the oxygens. (c) The uniform $\vec{h}_{\rm D}$ should be perpendicular to $\vec{S}^{\rm AFM}$ and $\vec{D}$.}
\vskip -0.4cm
\end{figure}

For simplicity, all spins in the AFM and FM side are assumed to be collinear. However, because of the different easy magnetic axes or planes for different materials, \emph{in general the NN spins are noncollinear at the FM/G-AFM interface} (Fig.~1(b)).  There are two vectors that are staggered: (1) the AFM interface spins $\vec{S}^{\rm AFM}_i$ given by ($-1$)$^i\vec{S}^{\rm A}$, and (2) the $\vec{D}_{ij}$ vectors across the interface given by ($-1$)$^i\vec{D}$, where $i$ denotes the site (or bond) sequence at the (001) interface. Combining these two staggered components $\vec{D}_{ij}$ and $\vec{S}^{\rm AFM}_i$, it is straightforward to obtain a \emph{uniform} DM effect at the interface:
\begin{equation}
H^{\rm interface}_{\rm DM}=\sum_{<ij>}\vec{D}_{ij}\cdot(\vec{S}^{\rm FM}_{i}\times\vec{S}^{\rm AFM}_{j})=-{\vec{h}_{\rm D}}\cdot{\sum_i\vec{S}^{\rm FM}_i},
\end{equation}
where $\vec{S}^{\rm FM}$ denotes the spin at the FM side and $i$ and $j$ only sums over the interface. $\vec{h}_{\rm D}$ is the effective magnetic field that points into the direction $\vec{D}\times\vec{S}^{\rm A}$ (Figs.~1(b,c)).
%(or its equivalent directions depending on the combinations 
%of $\vec{D}_{i,j}$ and $\vec{S}^{\rm AFM}_j$ used). 
Note that $\vec{h}_{\rm D}$ is uniform and independent of the FM spins' direction. The combination of $\vec{D}_{i,j}$ and $\vec{S}^{\rm AFM}_j$, namely $\vec{h}_{\rm D}$, can be fixed by the field-cooling process and then assumed to remain frozen at low $T$ during the hysteresis loop measurement \cite{note0}. Thus, this provides a bias field caused by the DM interaction which can produce a EB at interfaces of FM/G-AFM perovskites.

\textit{FE driven EB.} In the previous discussion, the second term (DM interaction) of Eq.(1) was proposed as the microscopic origin of EB in generic FM/G-AFM oxide heterostructures. However, the first term (SE) can also contribute to the EB {\it if} multiferroic materials are involved in the heterostructure. In ferroelectric (FE) materials, spontaneous relative displacements between cations and anions induce an electric polarization. Consider the oxygen positions at the interface shown in Fig.~2(a): in addition to the previously mentioned staggered displacements, that do not induce a finite $\vec{P}$, in some multiferroic materials their FE properties can be assumed to be caused by additional displacements of NN oxygens that should all be along the same direction to avoid a global cancellation. Therefore, \emph{the bond-angles at the interface can become asymmetric by the simultaneous consideration of these two displacement modes}. Since the SE coupling magnitude is dependent on the bond-angle \cite{Zhou:Prb}, these modulated bond-angles induce an interfacial SE coupling $J$ that is also staggered, with values that are denoted here as $J_{\rm L}$ and $J_{\rm S}$. Once again, as with the DM-driven EB, two staggered effects (alternating SE couplings at the interface, and alternating spin orientations on the AFM side of the interface) compensate each other. By this procedure, it is straightforward to obtain an additional uniform effective field at the interface:
\begin{equation}
H^{\rm interface}_{\rm SE}=\sum_{<ij>}J_{i,j}\vec{S}^{\rm FM}_{i}\cdot\vec{S}^{\rm AFM}_{j}=-\vec{h}_{\rm J}\cdot{\sum_i\vec{S}^{\rm FM}_i}.
\end{equation}
Here, $\vec{h}_{\rm J}=-\delta_{\rm J}\vec{S}^{\rm A}$ is the effective magnetic field, where $\delta_{\rm J}=(J_{\rm L}-J_{\rm S})/2$. 
%When there are many FE domains in a large region, the average $\vec{h}_{\rm J}$ most likely cancels. 
When an electric field is applied parallel to the interface to change the uniform polarization $\vec{P}$, the $\vec{h}_{\rm J}$ will change simultaneously, namely it is an \emph{electric-field-controllable} EB which is potentially important to design multiferroics devices.

\emph{Both $\vec{h}_{\rm J}$ and $\vec{h}_{\rm D}$ may have components parallel to the measuring field, although they 
are perpendicular to each other}. Experimentally, by varying the electric-field direction, estimations for the components of $\vec{h}_{\rm J}$ and $\vec{h}_{\rm D}$ can be obtained separately, since $\vec{h}_{\rm D}$ is almost independent of the FE $\vec{P}$, in a first-order approximation \cite{Sergienko:Prb}.

\begin{figure}
\centerline{\includegraphics[width=0.48\textwidth]{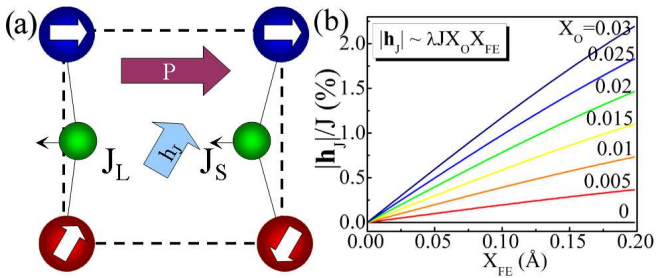}}
\caption{(Color online) (a) FE-polarization-driven asymmetric bond-angles and modulated normal SE at the interface. A switch of the FE polarization from left to right will also switch $\vec{h}_{\rm J}$. (b) The estimated $|{\bf h}_{\rm J}|$ as a function of $X_{\rm FE}$ for different values of $X_{\rm O}$ (from $0$ to $0.03$ \AA). The lattice constant is assumed to be $4$ \AA{} and $J_{i,j}$ is in proportional to $\cos^4(\theta_{i,j})$ where $\theta$ is the bond angle \cite{Zhou:Prb}. All displacements (in units of \AA{}) are assumed to be coplanar for simplicity.}
\vskip -0.4cm
\end{figure}

\textit{Discussion}. The basic physical picture related to the proposed DM- and FE-driven EB appears clear, but there are several practical issues that should be addressed.

First, in the derivations above, both mechanisms are independent of the details of the FM spins. Therefore, both mechanisms should be valid for a variety of FM materials such as perovskites \cite{Choi:Apl} or metallic alloys \cite{Chu:Nm}. The only condition needed is that the oxygen octahedra of the interfacial AFM cations must be complete, i.e. oxygens must bridge the two materials at the interface.

Also, our model should be robust against other tilting modes. For a general $a^{\alpha}b^{\beta}c^*$ mode, the NN $\vec{D}_{ij}$'s at the interface are not uniform as long as $\alpha$ and $\beta$ are not both simultaneously zero. If this is the case, a net $\vec{h}_{D}$ is still induced, with direction and value varying with the mode. For the tilting mode which only rotates along the [001] axis (the $a^0a^0c^+$ mode in the perfect tetragonal lattice), the DM contribution at the (001) interface is zero. In this case, $\vec{h}_{\rm J}$ will also be zero since the bond-angles are uniform. However, there is evidence that many perovskite films are not perfectly tetragonal \cite{note1}.
% Therefore, distortions of the M-O-M bonds at interfaces can exist. Of course, more careful studies should be performed to check the tilting modes/amplitudes and the spin configurations around the interface.

In addition, since the DM coupling is very weak (particularly in nearly tetragonal thin films), it is necessary to check whether the EB that it generates is compatible in magnitude with the experimental EBs. Considering $\vec{h}_{\rm D}$ to be only effective at the interface while the external magnetic field is applied on all FM spins, the maximum EB (when the measuring field is collinear with $\vec{h}_{\rm D}$) can be estimated as: $h_{\rm EB}$$\approx$$|\vec{h}_{\rm D}|/d$=$H^{\rm interface}_{\rm DM}/(dm)$, where $d$ is the FM material thickness in unit cells, and $m$ is the magnetic moment of the FM cation. In a first-order approximation, $|\vec{D}_{i,j}|$ is proportional to the oxygen displacement $X_{\rm O}$: $H^{\rm interface}_{\rm DM}$$\approx$$\gamma X_{\rm O}$, with $\gamma$ the DM coefficient roughly estimated as $1$ meV/\AA{} \cite{Sergienko:Prb}. A tiny distortion of the M$_1$-O-M$_2$ bond across the interface, as small as a $1^\circ$ bend \cite{note2}, can result in $H^{\rm interface}_{\rm DM}$$\approx$$0.0175$ meV if the lattice constant is $4$ \AA{}, indeed very weak compared with $J$ which is usually larger than $10$ meV for perovskites. Assuming typical values $d$=$10$ and $m$=$3$ Bohr magnetons, the DM driven EB is $100$ Oe which is of the same order of magnitude as the experimentally measured EBs in perovskite heterostructures \cite{Choi:Apl}.

For perovskite heterostructures involving multiferroics, both $\vec{h}_{\rm D}$ and $\vec{h}_{\rm J}$ 
should be considered. The estimated $|\vec{h}_{\rm J}|$ vs. the FE oxygen displacement ($X_{\rm FE}$, which is proportional to the in-plane projection of $\vec{P}$) and parametric with $X_{\rm O}$ are 
shown in Fig.~2(b). 
When both $X_{\rm O}$ and $X_{\rm FE}$ are small, $|\vec{h}_{\rm J}|$ 
behaves approximately as $\lambda JX_{\rm O}X_{\rm FE}$ ($\lambda$$\approx$$3.7$-$4.0$). Thus, 
$|\vec{h}_{\rm J}|/|\vec{h}_{\rm D}|$ 
is estimated to be $\lambda JX_{\rm FE}/\gamma$,
which may be larger than $1$ in BFO.

It is also important to analyze if the DM- and FE-driven EB mechanisms 
are robust against roughness, which often is appreciable at interfaces, 
although recent experimental progress in thin-films substantially reduces
this extrinsic effect. Since there are several uncertain factors controlling 
the interfacial roughness, it is difficult to reach robust conclusions from 
the theoretical perspective. For this reason 
a simplified analysis will be given here, 
by assuming that the FM and AFM cations can be mixed near the interface 
but they not diffuse into inner regions, as shown in Fig.~3. If the G-AFM spin order 
is stable enough and there are no crystal defects at the interface, the $\vec{D}$ 
vectors across the interface alway change simultaneously with the corresponding 
spins in the G-AFM side following the roughness. In other words, 
the roughness geometry would not change the combination of $\vec{D}_{ij}$ 
and G-AFM spin vectors at the interface. Even though the AFM spins can 
be canted at the roughened regions, this may decrease but will not cancel 
the global $\vec{h}_{\rm D}$, as long as there are no 
separated $180^\circ$ magnetic domains or ferroelastic walls. Similarly, 
it can be shown that the FE driven mechanism will not be canceled by
roughness either. Therefore, both the DM- and FE-driven mechanisms for EB 
should in principle work, even in the presence of weak interface roughness.

\begin{figure}
\centerline{\includegraphics[width=240pt]{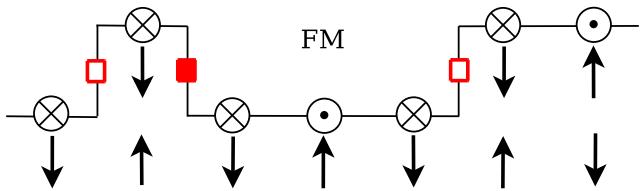}}
\caption{(Color online) Sketch of atomic-scale interface roughness. 
Only the ideal G-AFM spin order is shown by arrows. The alternation of the 
$\vec{D}_{ij}$ vectors across the interface 
are shown as in- and out-arrows. In addition, in the roughened case, the $D$ vectors 
of the (100) and (010) bonds (open/full squares) will also be active for the EB.}
\vskip -0.4cm
\end{figure}

Note also that the DM- and FE-driven EB are 
anisotropic (related to the crystal direction).
%In the LSMO/BFO heterostructures,  the EB is indeed anisotropic in-plane with the optimized directions along the [100]/[010] axes \cite{Huijben:un}. 
Ideally, if the measuring field is 
applied perpendicular to $\vec{h}_{\rm D}$ (if no multiferroics are 
involved), there would be no EB. A possible example 
is the case of LaMnO$_3$/SMO superlattices, in which no EB has been 
observed using an in-plane measuring field since all spins are almost in-plane 
collinear (thus $\vec{h}_{\rm D}$ is out-of-plane) \cite{Bhattacharya:Prl}.

Finally, we remark that we have tested our argumentations using numerical techniques
on a heterostructure \cite{note1}, and a robust EB in the hysteresis loop 
was obtained by considering our two mechanisms.
%An easy magnetic axis for the AFM material is needed to stabilize the AFM spins' direction. 
In the simulation, spin canting effects (that can originate from 
finite-$T$ fluctuations, exchange couplings at the interface,  
or magnetic field reorientation) are included, but they are not found to affect our 
results qualitatively.
%Quantitative simulations will be presented in the near future when accurate parameters are known for particular materials.

\textit{Conclusions.}
Here it was proposed that both the Dzyaloshinskii-Moriya interaction and the standard 
superexchange 
(the latter active only when multiferroic materials  
that can be controlled by electric fields are involved) 
could induce the exchange bias phenomenon 
at FM/G-AFM perovskite oxides interfaces, even when the antiferromagnetic spins 
are compensated. The common precondition for the existence of 
these two mechanisms is the presence of oxygen 
octahedral tiltings at the interface. Our model highlights the interactions between 
magnetism and lattice distortions, and proposes mechanisms to understand 
the exchange bias in FM/G-AFM oxides heterostructures.

We thank R. Ramesh, P. Yu, L.W. Martin, and M. Huijben for providing experimental 
results before publication and fruitful discussions. We also thank C. Ederer, J.-S. Zhou, 
O. Chmaissem, S. Okamoto, and J. Nogu\'{e}s for helpful discussions. 
Work supported by the NSF (DMR-0706020) and the Division of 
Materials Science and Engineering, U.S. DOE, under contract with UT-Battelle, LLC. K.Y. and S.P. were supported by the European Research Council under the EU 7$^{th}$ Framework Programme (FP7/2007-2013)/ERC grant agreement n. 203523. S.Y. was supported by CREST-JST. J.M.L was supported by the 973 Projects of China (2009CB929501) NSFC (85032002).


\begin{thebibliography}{bib}
\bibitem{note0} Due to anisotropies, the AFM spins are assumed to be 
almost unaffected during the measuring process. The possible tiny canting following the measuring field will not affect our mechanisms. Also, due to the energy barriers between the lattice 
distortion directions, $\vec{D}_{i,j}$ can also be assumed to be frozen at low $T$.
\bibitem{note2}
The bond-angles are different along different axes. Here only the bond along the [001] 
axis contributes to the EB.
\end{thebibliography}
\end{document}